\documentclass{svproc}
\usepackage{url}

\usepackage{graphicx}
\usepackage{bm}
\usepackage{hyperref}
\usepackage{multicol}

\begin{document}
\mainmatter              
\title{Freeze-out of Strange Hadron in pp, pPb and PbPb Collisions at LHC Energies}

\titlerunning{Strange Hadron Production}

\author{Kapil Saraswat $^{1}$ \and Prashant Shukla $^{2,3}$ \and Vineet Kumar $^{2}$ 
\and Venktesh Singh $^{1}$}

\authorrunning{Kapil Saraswat et al.} 

\institute{$^{1}$ Department of Physics, Banaras Hindu University, Varanasi 221005, India \\ 
$^{2}$ Nuclear Physics Division, Bhabha Atomic Research Centre, Mumbai 400085, India \\
$^{3}$ Homi Bhabha National Institute, Anushakti Nagar, Mumbai 400094, India}

\maketitle              

\begin{abstract}
{\footnotesize
In this article, we will present a systematic analysis of transverse momentum spectra 
of the strange hadron in different multiplicity events produced in pp collision at 
$\sqrt{s}$ = 7 TeV, pPb collision at $\sqrt{s_{NN}}$ = 5.02 TeV and PbPb collision at 
$\sqrt{s_{NN}}$ = 2.76 TeV. The differential freeze out scenario of strange hadron 
$K^{0}_{s}$ assumed while analyzing the data using a Tsallis distribution which is
 modified to include transverse flow. The $p_{T}$ distributions of strange hadron 
in different systems are characterized in terms of the parameters namely, Tsallis 
temperature ($T$), power ($n$) and average transverse flow velocity ($\beta$).}
\end{abstract}
\section{Introduction}

{\footnotesize The transverse momentum $p_{T}$ spectra of the hadrons measured in pp , 
pPb and PbPb collisions reflect the condition of the system at the time of freeze-out. 
The $p_{T}$ spectra of the hadrons are very useful tools to study the particle production 
mechanisms, thermalization and collective effects and at high $p_{T}$ probe jet quenching 
effects.}

\section{Modified Tsallis Distribution Function}
{\footnotesize The transverse momentum spectra of hadrons can be described using the 
modified Tsallis distribution \cite{Khandai:2013fwa}. The modified Tsallis function 
is given by}
\begin{scriptsize}
\begin{equation}
E \frac{d^{3}N}{dp^{3}} = C_{n} \Bigg[\exp\Bigg(\frac{- \gamma ~ \beta ~ p_{T}}{n ~ T}\Bigg) + 
\frac{\gamma ~ m_{T}}{ n ~ T}\Bigg]^{-n}~.
\label{pbpb_equation1one}
\end{equation}
\end{scriptsize}

{\footnotesize Here $C_{n}$ is the normalization constant, $m_{T} (\sqrt{p^{2}_{T} + m^{2}})$ 
is the transverse mass, $\gamma = 1/\sqrt{ 1 - \beta^{2}}$, $\beta$ is the average transverse 
velocity of the system and $T$ is the temperature. Phenomenological studies suggest that, 
for quark-quark point scattering, $n~ \sim$~4 and when  multiple scattering centers are 
involved $n$ grows larger.}

\section{Results and Discussions}

\begin{figure}[!htb]
\centering
\begin{minipage}{.50\textwidth}
\centering
\includegraphics[width=0.90\linewidth, height=0.26\textheight]{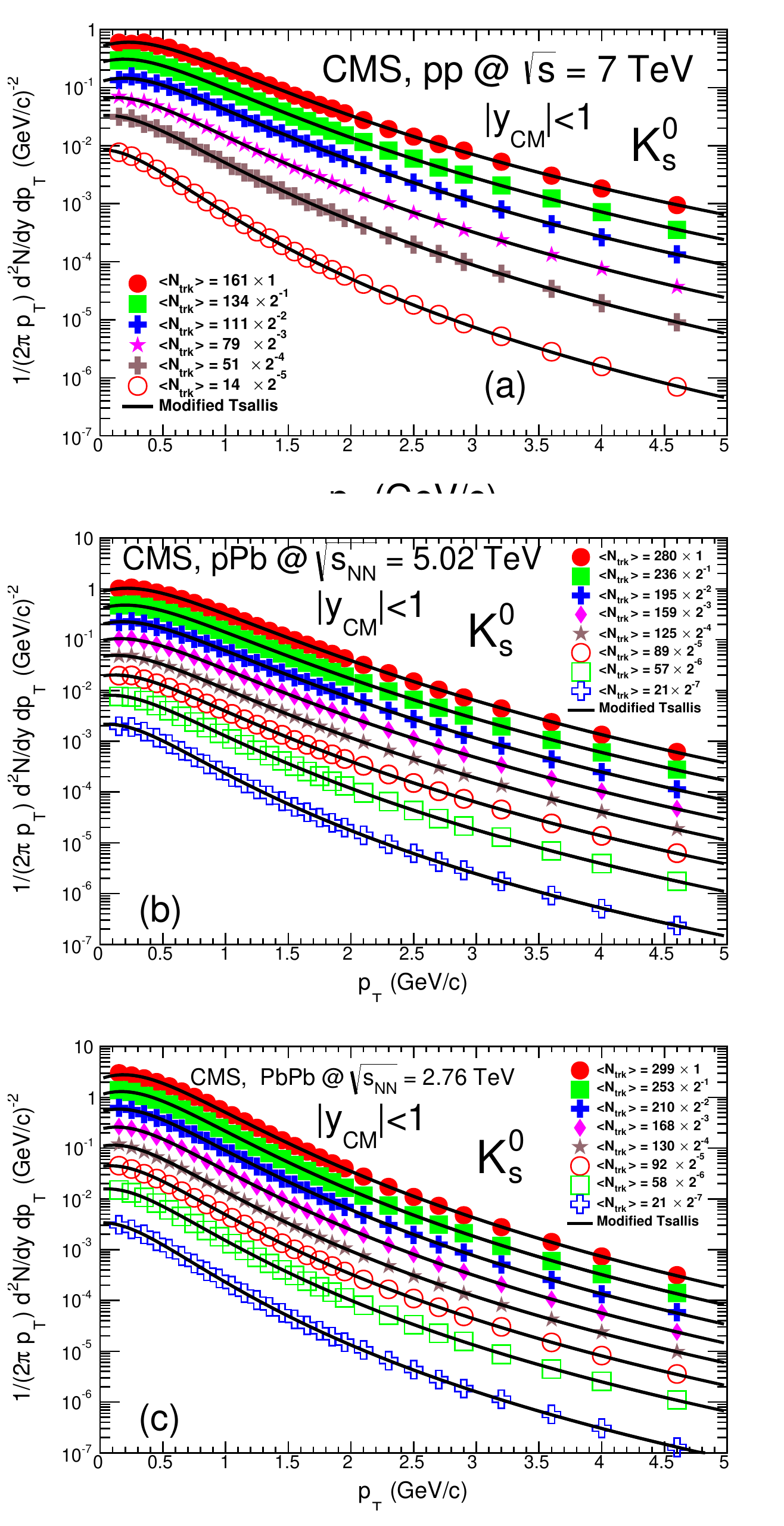}
\caption{{\footnotesize The invariant yield of $K^{0}_{s}$ as a function of pT for pp, pPb 
and PbPb collisions.}}
\label{figure1_fitting}
\end{minipage}%
\begin{minipage}{0.50\textwidth}
\centering
\includegraphics[width=0.90\linewidth, height=0.26\textheight]{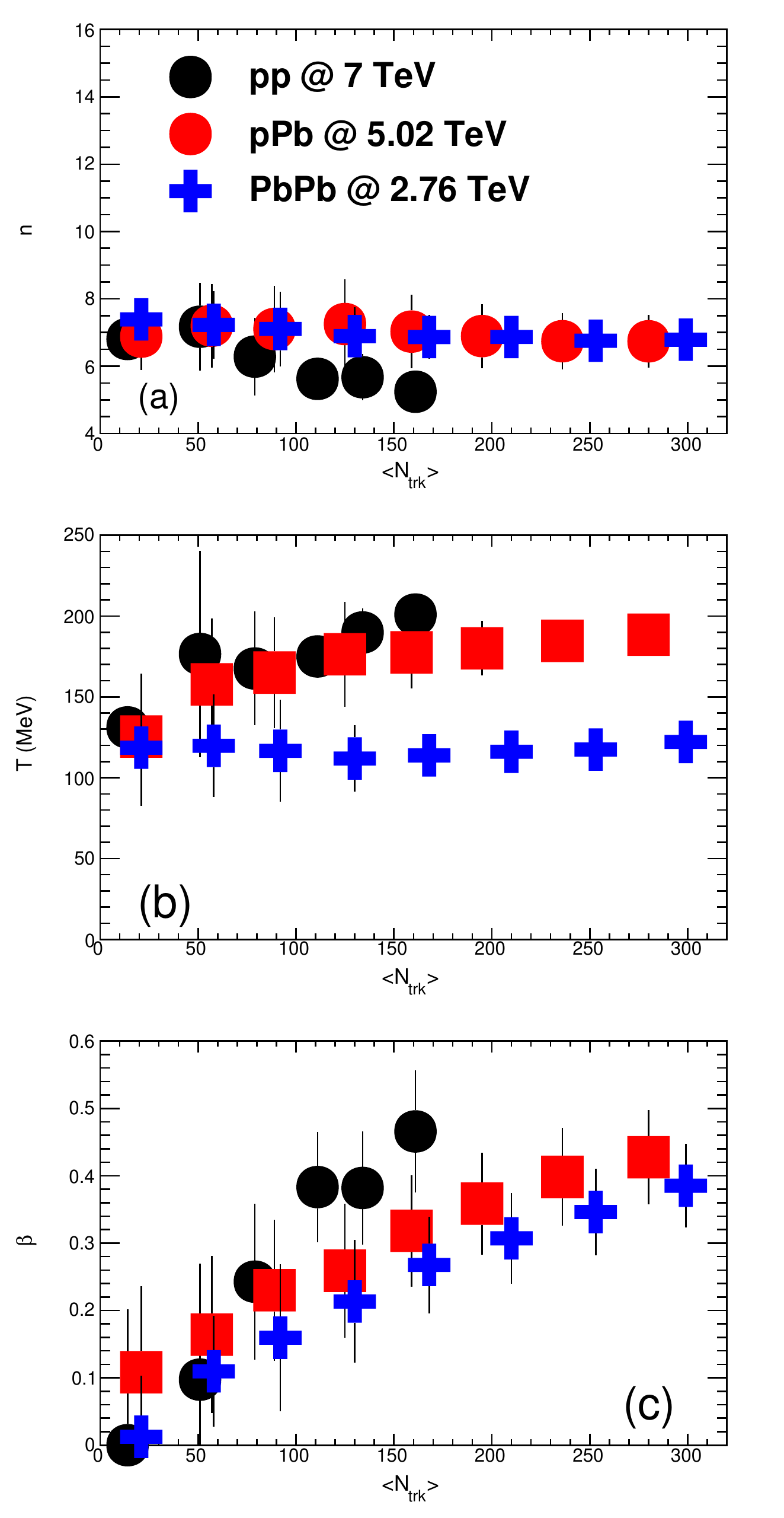}
\caption{{\footnotesize $n, T$ and $\beta$ for $K^{0}_{s}$ as a function of $<N_{\rm{trk}}>$ 
in pp, pPb and PbPb collisions}}
\label{figure2_fitting_parameter}
\end{minipage}
\end{figure}

{\footnotesize Figure \ref{figure1_fitting}  shows the invariant  yields of the strange 
hadron $K^{0}_{s}$ $\Big($ (a) pp collision at $\sqrt{s}$ = 7 TeV, (b) pPb collision at 
$\sqrt{s_{NN}}$ = 5.02 TeV and (c) PbPb collision at $\sqrt{s_{NN}}$ = 2.76 TeV $\Big)$ 
as a function of the $p_{T}$ measured by the CMS experiment \cite{Khachatryan:2016yru} 
in the mid rapidity $|y_{\rm{CM}}|<$ 1. The invariant yields are given for different 
multiplicity classes. The solid curves are the modified Tsallis distribution fitted 
to hadron $p_{T}$ spectra. The modified Tsallis gives good description of the data.}

{\footnotesize Figure \ref{figure2_fitting_parameter} (a) shows the Tsallis parameter 
$n$ as a function of the efficiency corrected average track multiplicity $<N_{\rm{trk}}>$ 
in the pp collision at $\sqrt{s}$ = 7 TeV, pPb collision at $\sqrt{s_{NN}}$ = 5.02 TeV 
and PbPb collision at $\sqrt{s_{NN}}$ = 2.76 TeV.  The parameters $T$ and $\beta$ are 
shown in the panels (b) and (c) respectively. The value of $n$ decreases with multiplicity 
in pp collision but in pPb and PbPb collisions it remains similar. The value of the Tsallis 
temperature $T$ increases with the multiplicity in pp and pPb collisions but it varies 
little with the multiplicity. The transverse flow $\beta$ increases with the multiplicity 
in all three systems.}

\section{Conclusion}
{\footnotesize We have studied the $p_{T}$ spectra of $K^{0}_{s}$ in three systems pp, 
pPb and PbPb. The value of the parameter $n$ has a little variation with multiplicity 
implying that the degree of thermalization remains similar for the events of different 
multiplicity classes in all three systems. The Tsallis temperature $T$ increases with 
the multiplicity for pp and pPb systems but it does not show a noticeable change with 
the event multiplicity for PbPb systems. The PbPb system has smaller temperature and 
transverse flow  as compared to the pp and pPb systems.}

%

\end{document}